\documentclass{ws-ijmpa}

\begin{document}

\markboth{Xiaoyan SHEN}{Recent $J/\psi$ Results from BESII}

\catchline{}{}{}{}{}

\title{Recent $J/\psi$ Results from BESII}

\author{Xiaoyan SHEN (for the BES Collaboration)}
\address{Institute of High Energy Physics, Chinese Academy of Science, \\
Beijing 100049, People's Republic of China}


\maketitle


\begin{abstract}
The studies on the multi-quark candidates,
light scalar mesons and excited baryon states at BES are presented,
based on $5.8 \times 10^7$ $J/\psi$ data collected with BESII detector.
The measurements of some $J/\psi$ and $\eta_c$ decays are presented too.
We also report the searches for the lepton flavor violation and
pentaquark states in $J/\psi$ decays.
\end{abstract}

\keywords{multiquark state; $J/\psi$ decay; scalar meson, 
branching ratio; baryon}

\section{Introduction}


BES, as described in ref. \cite{bes}, is a large general purpose solenoidal 
detector at the Beijing Electron Positron Collider (BEPC). 
Since 1998, the BES detector has been upgraded to BESII. 
The $5.8 \times 10^7$ $J/\psi$ events have been accumulated with BESII
since then, which provides a good laboratory for the study of the
non-$q \bar q$ states and hadron spectroscopy.

\section{Study of the multiquark candidates}

\subsection{Near $p \bar p$ threshold enhancement in $J/\psi \to
\gamma p \bar p$}

\begin{figure}[htpb]
\centerline{\psfig{file=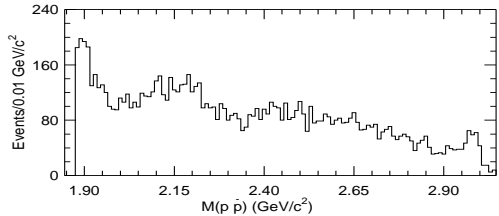,width=6.5cm,height=2.8cm}}
 \caption{The $ p \bar p$ invariant mass distribution in
$J/\psi\to \gamma p \bar p$ decays.}
    \label{2pg_data_ihep}
\end{figure}

 There is an accumulation of evidence for anomalous behavior in the 
$p \bar p$ system near $2m_p$ mass threshold. We analyze 
$J/\psi \to \gamma p \bar p$ with BESII $J/\psi$ data \cite{gppb}. 
Fig. ~\ref{2pg_data_ihep} shows the $p \bar p$ invariant mass
distribution for selected events. Except for
a peak near $M_{p \bar p}=2.98$~GeV/$c^2$ that is consistent
in mass, width, and yield with expectations for $J/\psi\to\gamma\eta_c$,
$\eta_c\to p \bar p$~\cite{etac} and a broad enhancement around
$M_{p \bar p}\sim 2.2$~GeV/$c^2$, there is a narrow, low-mass peak
near the $p \bar p$  mass threshold.

The low mass region of the $p \bar p$ distribution is fitted by an
acceptance-weighted $S$-wave Breit-Wigner (BW) function and
$f_{\rm bkg}(\delta)$ which represent the low-mass enhancement and 
the background, respectively.
The fit yields $928\pm 57$ events
in the BW function with a peak mass of
$M=1859 ^{~+3}_{-10}$ $^{~+5}_{-25}$~MeV/$c^2$ and a full
width of $\Gamma < 30$~MeV/$c^2$ at a 90\% confidence
level (CL).

\subsection{The anomalous enhancements near the $m_p + M_{\bar \Lambda}$ and
$m_K + M_{\bar \Lambda}$ mass thresholds in $J/\psi \rightarrow p K^- \bar
\Lambda + c.c.$ decays}

\begin{figure}[htpb]
\centerline{\psfig{file=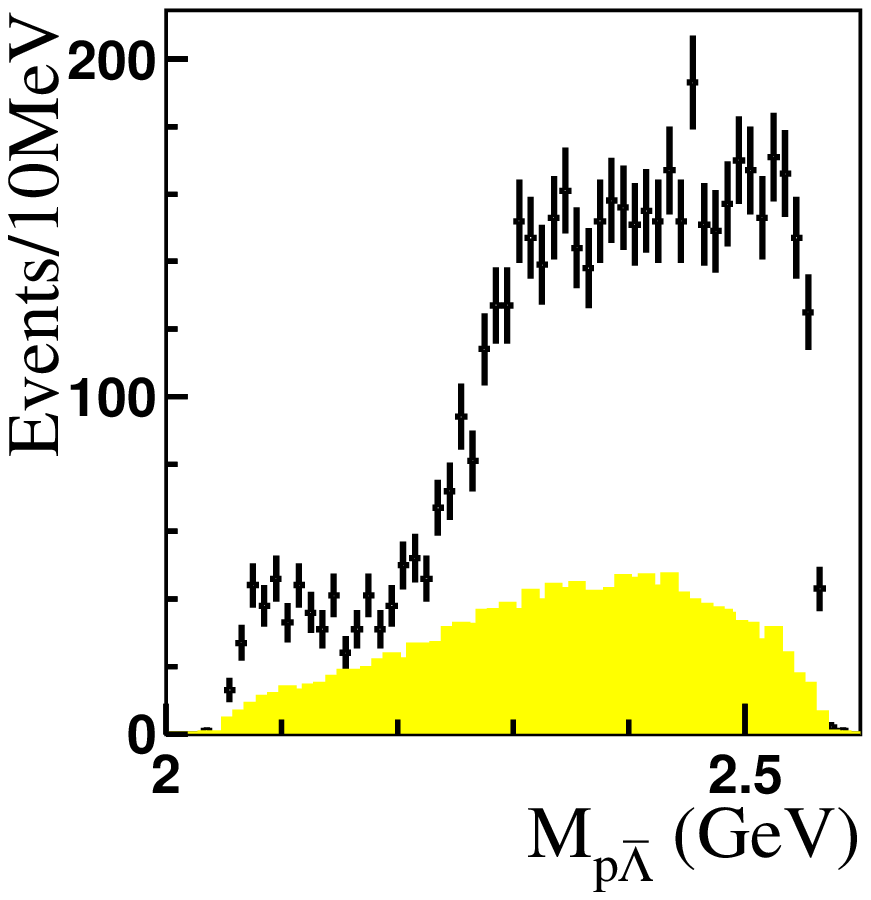,width=4.8cm,height=2.8cm}
  \psfig{file=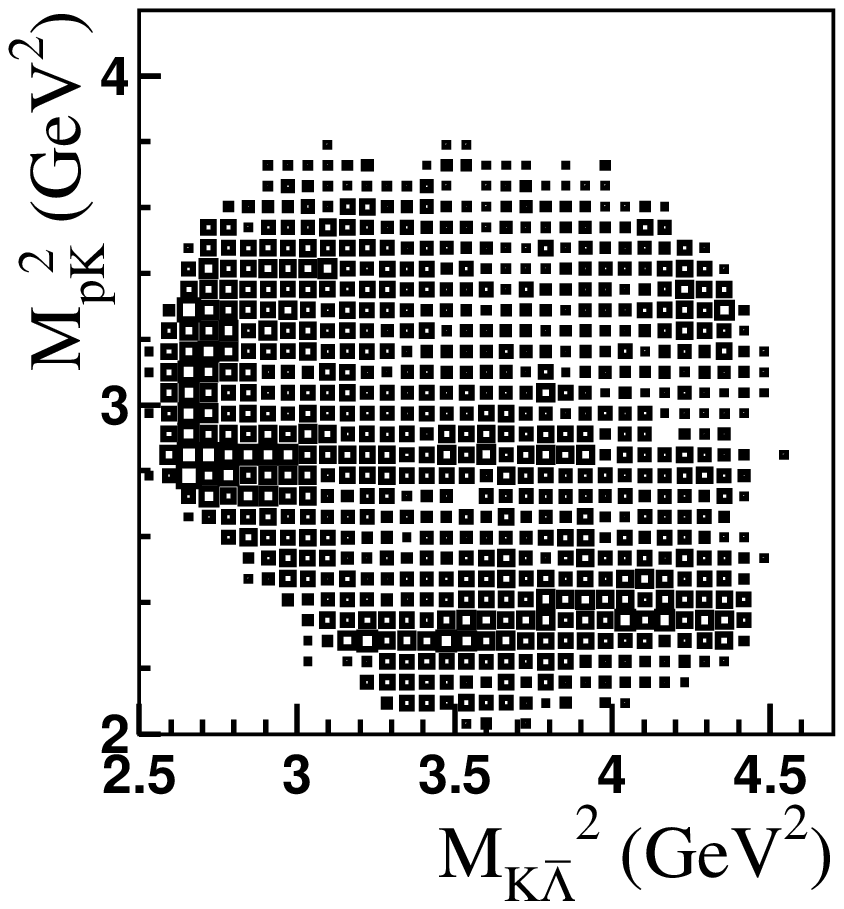,width=4.5cm,height=2.8cm}}
 \caption{Left: the points with error bars indicate the measured
            $p \bar \Lambda$ mass spectrum; the shaded histogram indicates
            phase space MC events (arbitrary normalization).
            Right: the Dalitz plot for the selected event sample.}
    \label{x208}
\end{figure}

The decay of $J/\psi \rightarrow pK^- \bar \Lambda + c.c$ are 
studied \cite{pkl}.
Fig.~\ref{x208}~(left) shows the $p \bar \Lambda$ invariant mass spectrum
for the selected events, where an enhancement is evident near the mass
threshold. The $pK^- \bar \Lambda$ Dalitz plot is shown in
Fig.~\ref{x208}~(right). In addition to bands for the well established
$\Lambda^*(1520)$ and $\Lambda^*(1690)$, there is a significant $N^*$ band
near the $K^- \bar{\Lambda}$ mass threshold, and a $p \bar{\Lambda}$ mass
enhancement in the right-upper part of the Dalitz plot, isolated from 
the $\Lambda^*$ and $N^*$ bands.

This enhancement can be fitted with an acceptance weighted S-wave Breit-Wigner
together with a function $f_{PS}(\delta)$ describing
the phase space contribution.
The fit gives a peak mass of $m=2075\pm 12$~MeV, a width
$\Gamma=90 \pm 35$~MeV and a branching ratio
$$BR(J/\psi \rightarrow K^-X) BR (X\rightarrow p \bar \Lambda)
 =(5.9\pm 1.4)\times 10^{-5}.$$

\vspace{-0.5cm}

\vspace{-0.5cm}
\begin{figure}[htpb]
\centerline{\psfig{file=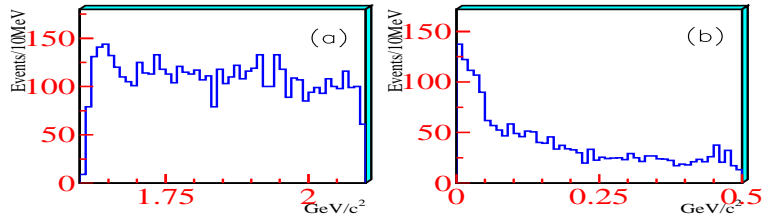,width=10.cm,height=2.8cm}}
 \caption{(a). The $m_{K^- \bar \Lambda}$ invariant mass spectrum
                      from $J/\psi \to pK^- \bar \Lambda$.
         (b). The $m_{K^- \bar \Lambda}-m_{K^-}-m_{\bar \Lambda}$
            distribution after the efficiency and phase
            space correction.}
        \label{kl}
\end{figure}

As mentioned above, the $pK^- \bar \Lambda$ Dalitz plot,
Fig.~\ref{x208} (right), shows a significant $N^*$ band near the
$K^- \bar{\Lambda}$ mass threshold. This band corresponds to
an enhancement near the $K^- \bar{\Lambda}$ mass threshold
in the one dimension projection of $m_{K^- \bar \Lambda}$,
shown in Fig. ~\ref{kl}(a). The $m_{K^- \bar \Lambda}
-m_{K^-}-m_{\bar \Lambda}$ distribution, Fig. ~\ref{kl}(b), 
after the efficiency and phase
space correction presents a more obvious peak at the $K^-
\bar{\Lambda}$ mass threshold.
                                                                                
The partial wave analysis results show that the mass and width of this
enhancement, $N_x$, are around 1500 - 1650 MeV and 70 - 110 MeV, respectively, 
the $J^P$ favors $1/2^-$
and the product branching ratio $Br(J/\psi \to p \bar{N_x}) \times
Br(\bar{N_x} \to K^- \bar \Lambda)$ is around $2 \times 10^{-4}$. The big product
branching ratio indicates a large coupling of $N_x$ to $K \Lambda$.

\section{Light scalar mesons}

\subsection{The $\sigma$ and $\kappa$}

There have been hot debates on the existence of $\sigma$ and $\kappa$.
The decay of $J/\psi \to \omega \pi ^+\pi ^-$, with the $\omega$ decaying 
to $\pi^+ \pi^- \pi^0$, is studied for $\sigma$ \cite{sigma}.

\begin{figure}[htbp]
\vspace{-0.5cm}
\centerline{\psfig{file=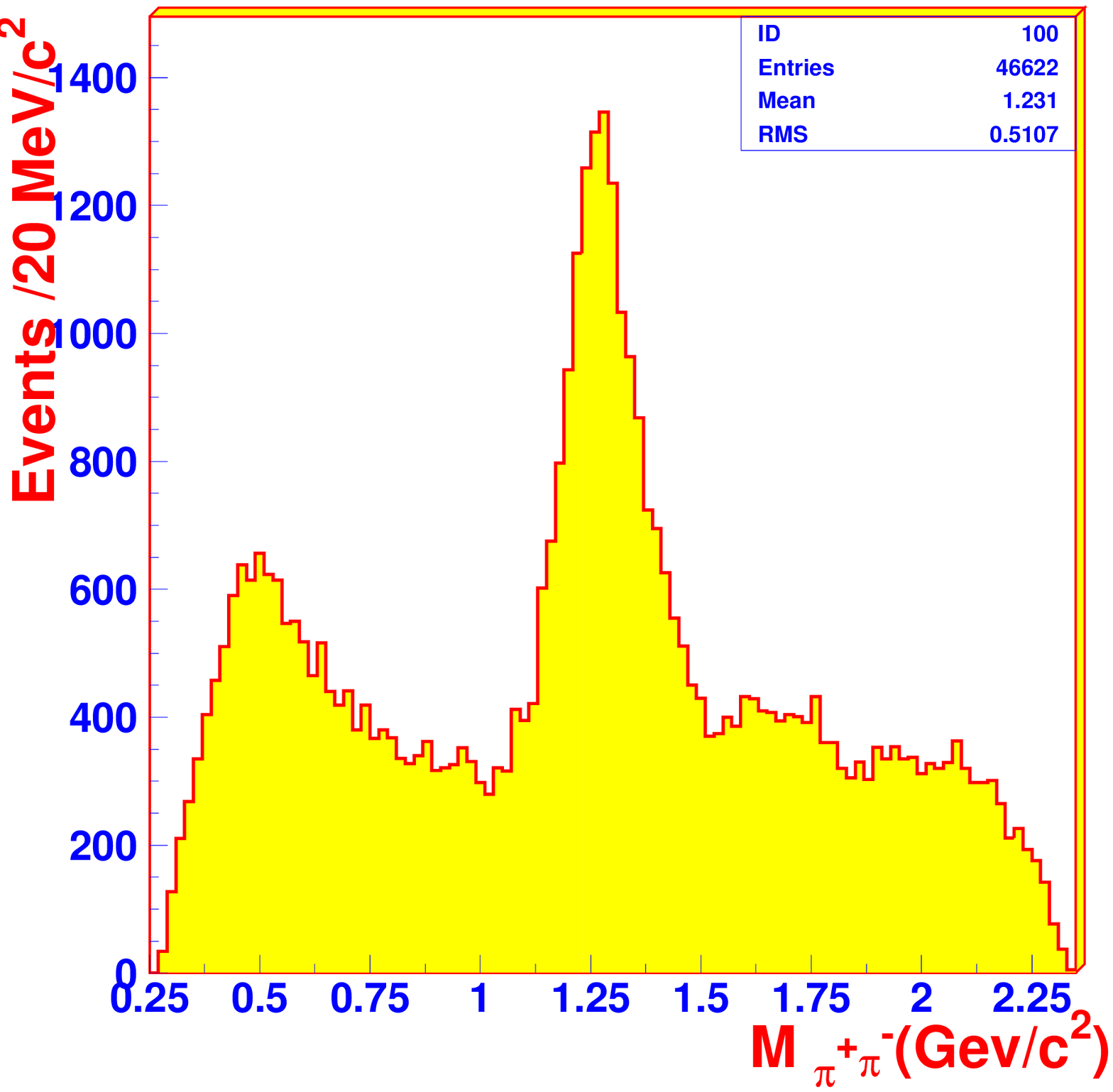,width=4.8cm,height=2.6cm}
  \psfig{file=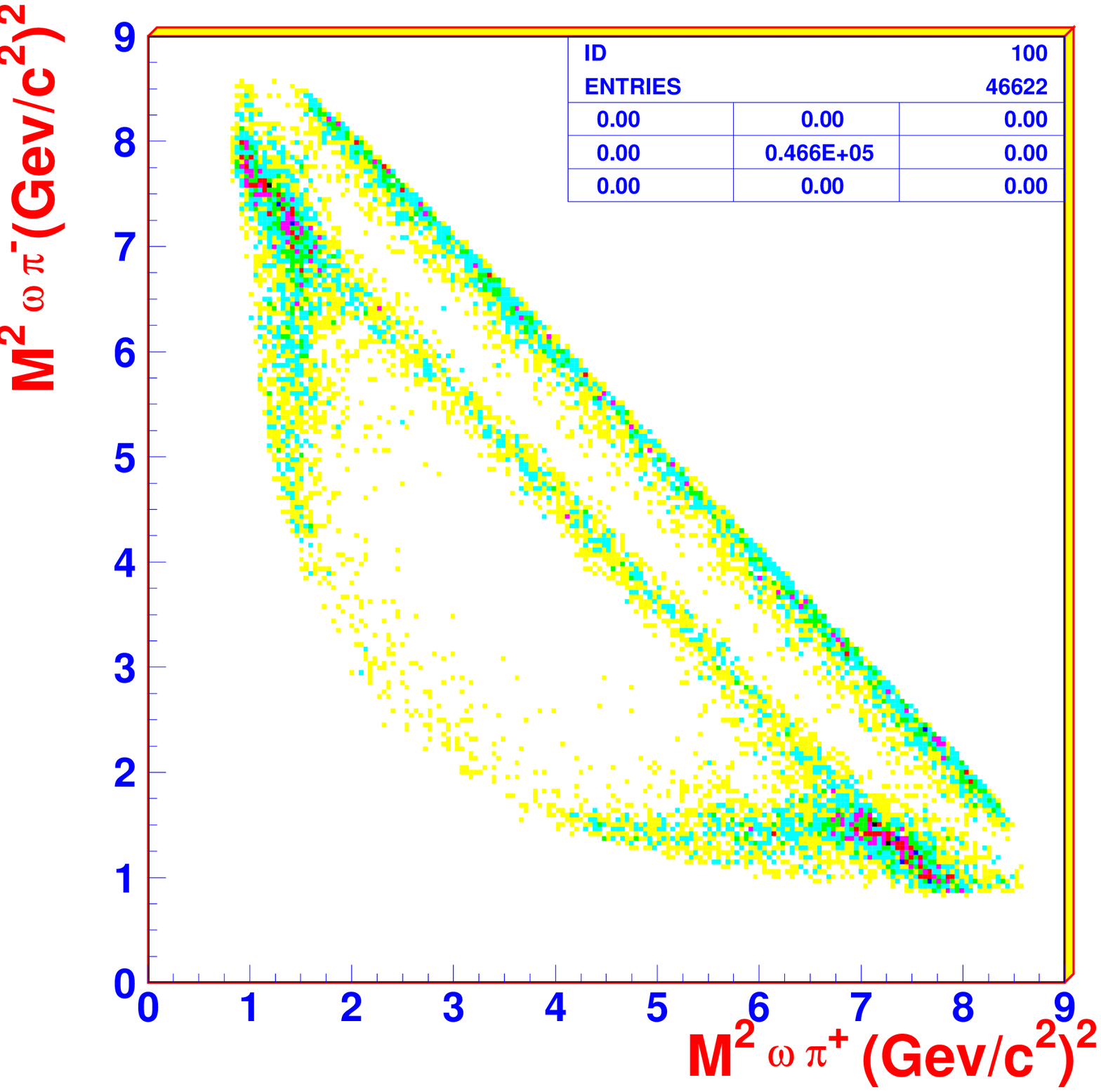,width=4.5cm,height=2.6cm}}
\caption{Left: distribution of $\pi ^+\pi ^- \pi ^0$ mass.
  Right: Dalitz plot.}
        \label{sigma}
\end{figure}

Fig. \ref{sigma} (left) shows the $\pi^+ \pi^-$ invariant mass spectrum 
recoiling against the $\omega$ for the selected $J/\psi \to \omega \pi^+
\pi^-$ events. The Dalitz plot of this channel is shown in 
Fig. \ref{sigma} (right). At low $\pi \pi $ masses, 
a broad enhancement which is due to the $\sigma$ 
pole is clearly seen. This peak is evident as a strong band along the 
upper right-hand edge of the Dalitz plot.

Two independent Partial wave analyses (PWA) have been performed on this 
channel. Different analysis methods and four 
parametrizations of the $\sigma$ amplitude give consistent results for 
the $\sigma$ pole. The average $\sigma$ pole position is determined to be
$(541 \pm 39) - $i$(252 \pm 42)$ MeV.

The $\kappa$ is studied from $J/\psi \to \bar{K^*}(892)K^+\pi^-$ and
$K^+K^-\pi^+\pi^-$ decays through partial wave analysis. Three
independent analyses have been performed and different parametrizations
of $\kappa$ pole are used. The preliminary results show the evidence
of $\kappa$ near the $K\pi$ threshold. Its pole position is around
$(760\sim840) -$i$(310\sim 420)$ MeV.


\subsection{$J/\psi\to \omega K^+K^-$}

\begin{figure}[htbp]
\centerline{\psfig{file=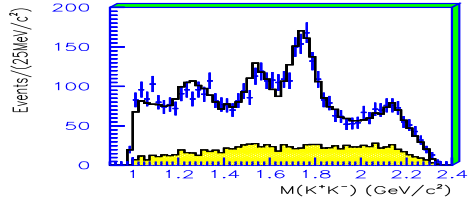,width=6.5cm,height=3.5cm}}
\caption{The $K^+ K^-$ invariant mass distribution from
$J/\psi \to \omega K^+K^-$ (crosses).
The full histogram shows the maximum likelihood fit and the shaded histogram
the background estimated from $\omega$ sidebands.}
\label{wkk}
\end{figure}

Fig. \ref{wkk} shows the $K^+K^-$ invariant mass distribution from
$J/\psi \to \omega K^+K^-$. The crosses are data and the shaded area 
indicates background events from the $\omega$ sideband estimation. 
A dominant feature of this channel is the structure around 1.74 GeV,
denoted as $f_0(1710)$. 
A partial wave analysis (PWA) shows that the $J^P$ of this structure favors
$0^+$ and the mass and width are optimized at $M = 1738 \pm 30$ MeV, 
$\Gamma = 125 \pm 20$ MeV. 
\vspace{-0.7cm}

In $J/\psi \to \omega \pi ^+ \pi ^-$~\cite{sigma}, there is no definite 
evidence for the presence of $f_0(1710)$. Therefore, we find 
at the 95\% confidence level
$$\frac {BR(f_0(1710) \to \pi \pi )}{BR(f_0(1710) \to K\bar K )} < 0.11.$$

\subsection{$J/\psi\to \phi \pi^+\pi^-$ and $J/\psi\to \phi K^+K^-$
(preliminary)}

\begin{figure}[htbp]
\centerline{\psfig{file=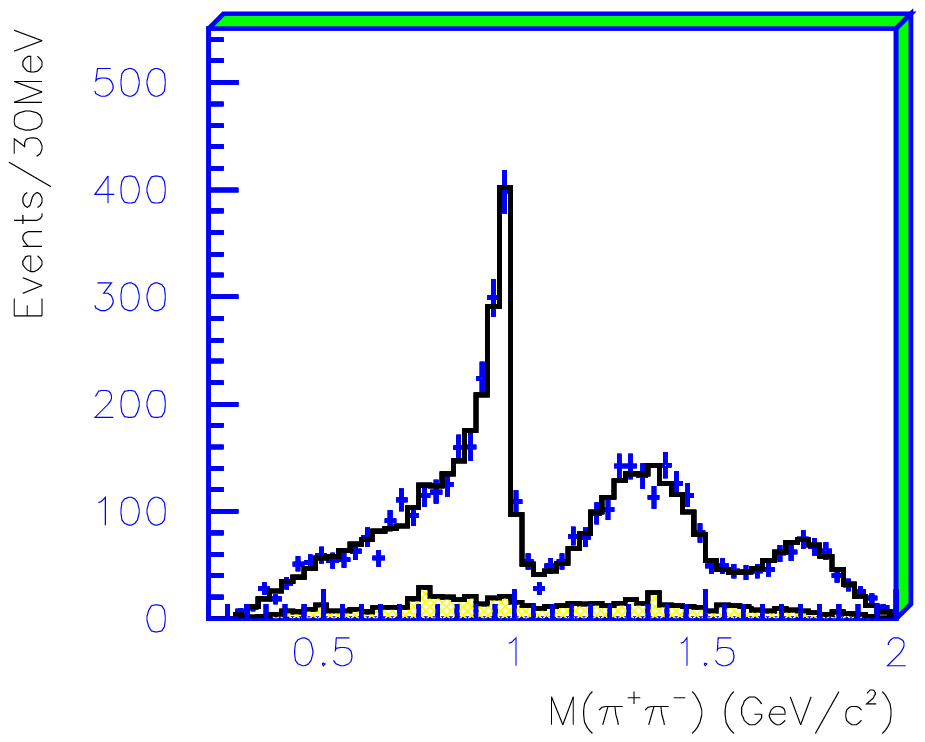,width=4.8cm,height=3.4cm}
  \psfig{file=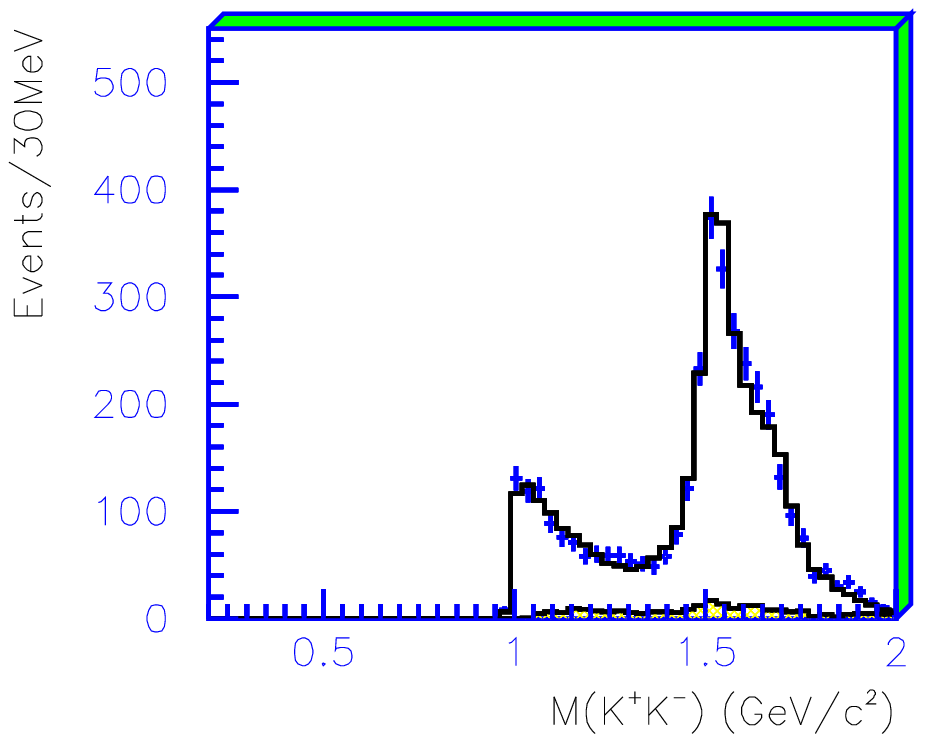,width=4.8cm,height=3.4cm}}
 \caption{Left: the mass spectrum of $\pi^+\pi^-$ in $J/\psi \to 
               \phi \pi^+\pi^-$.
          Right: the mass spectrum of $K^+K^-$ in $J/\psi \to
               \phi K^+ K^-$. Crosses are data and histograms are PWA
                fit projections.}
    \label{fppkk}
\end{figure}

Fig. \ref{fppkk} (a) and (b) show the $\pi^+\pi^-$ and $K^+K^-$ invariant 
mass distributions from $J/\psi \to \phi \pi^+\pi^-$ and $\phi K^+K^-$,
respectively. The shaded histograms correspond to the backgrounds estimated 
from $\phi$ sidebands.

The $\phi \pi ^+\pi ^-$ and $\phi K^+K^-$ data are fitted
simultaneously by using partial wave analysis, constraining resonance 
masses and widths to be the same in both sets of data.
The full histograms in Fig. \ref{fppkk} (left) and (right) show the maximum 
likelihood fit.
               
\vspace{-0.7cm}
                                                                 
The $f_0(980)$ is observed clearly in both sets of data.
The Flatt\' e form:
$$f=\frac {1}{M^2 - s - i(g_1\rho _{\pi \pi } +
g_2\rho _{K\bar K})}.$$
has been used to fit the $f_0(980)$ amplitude.
Here $\rho$ is Lorentz invariant phase space, $2k/\sqrt {s}$, $k$
refers to $\pi$ or $K$ momentum in the rest frame of the
resonance.
The present data offer the opportunity to determine the
parameters of $f_0(980)$ accurately:
$M = 965 \pm 8(sta) \pm 6(sys) $ MeV,
$g_1 = 165 \pm 10(sta) \pm 15(sys)$ MeV,
$g_2/g_1 = 4.21 \pm 0.25(sta) \pm 0.21(sys)$.

The $\phi \pi \pi$ data also exhibit a strong peak centered at
$M = 1335$ MeV. It may be fitted with $f_2(1270)$ and a dominant $0^+$ signal
made from $f_0(1370)$ interfering with a smaller $f_0(1500)$ component.
The Mass and width of $f_0(1370)$ are determined to be:
$M = 1350 \pm 50$ MeV and $\Gamma = 265 \pm 40$ MeV.

There is also a signal at around 1.79 GeV in $J/\psi \to \phi \pi^+ \pi^-$
with $M = 1790 ^{+40}_{-30}$ MeV and $\Gamma = 270 ^{+60}_{-30}$ MeV. The
spin 0 is prefered over spin 2.

\subsection{$J/\psi \to \gamma \pi\pi$ (preliminary)}

\begin{figure}[htbp]
\centerline{\psfig{file=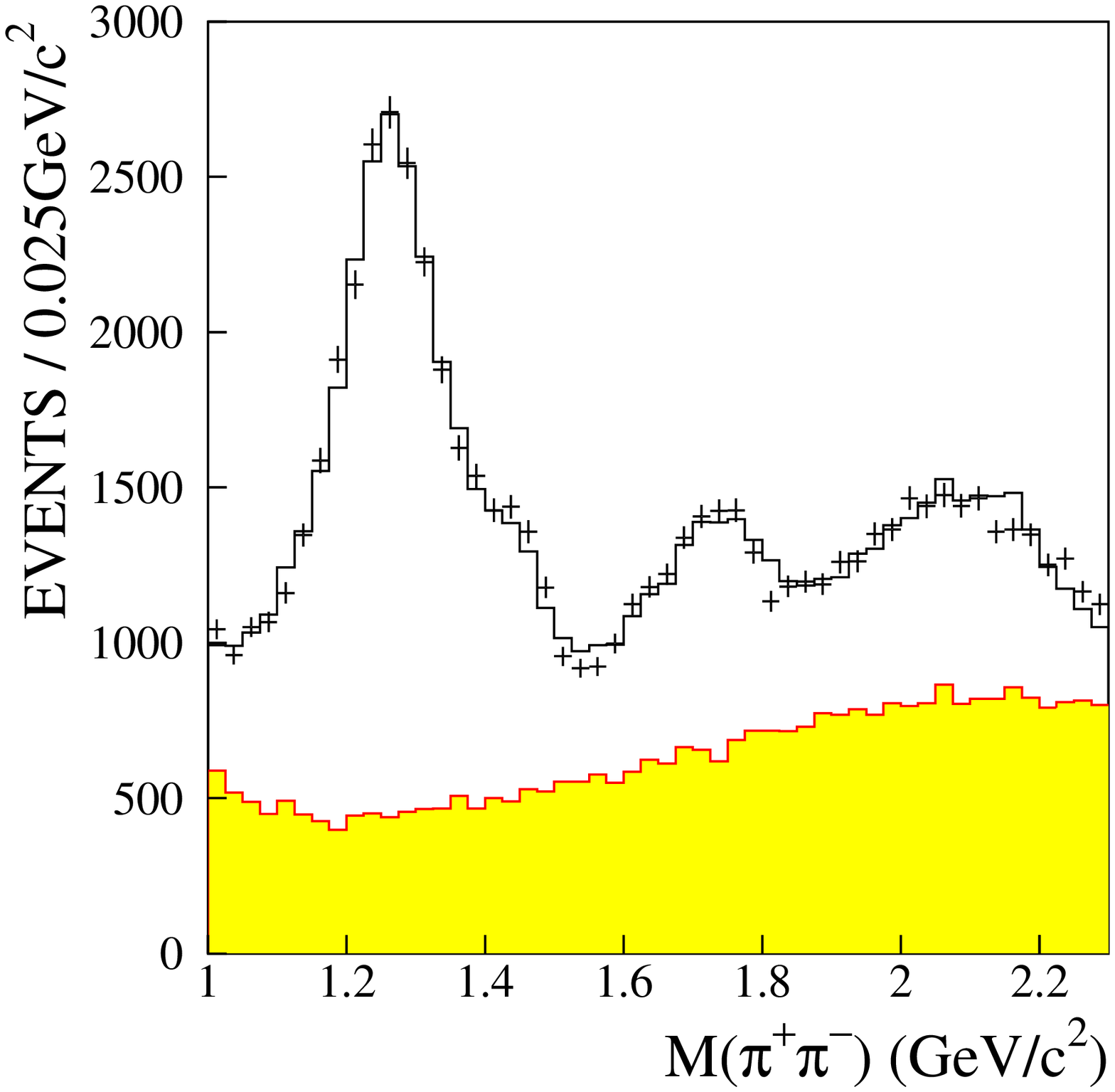,width=4.8cm,height=2.8cm}
  \psfig{file=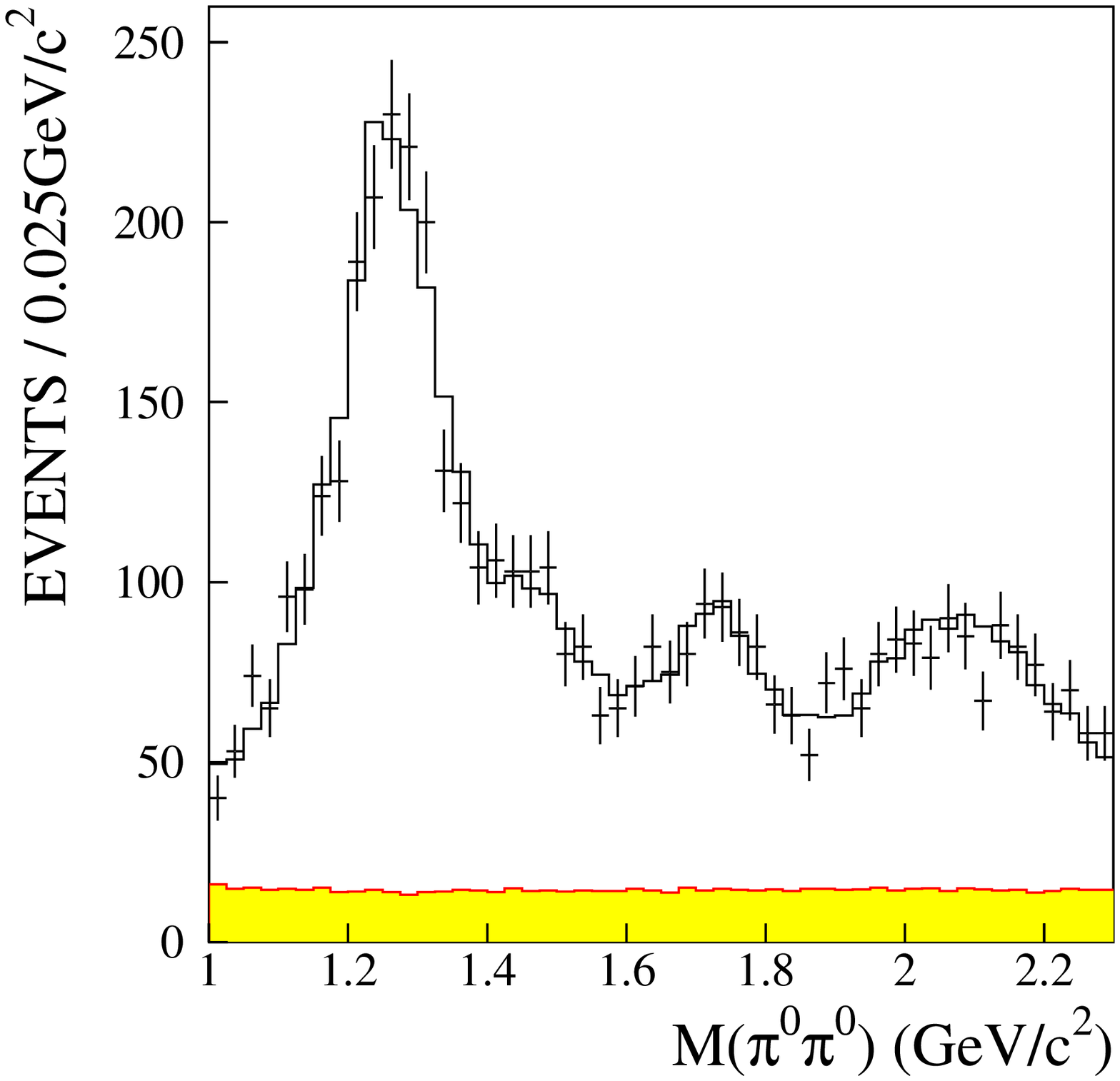,width=4.8cm,height=2.8cm}}
 \caption{Left: the mass spectrum of $\pi^+\pi^-$ in $J/\psi \to
               \gamma \pi^+\pi^-$.
          Right: the mass spectrum of $\pi^0 \pi^0$ in $J/\psi \to
               \gamma \pi^0 \pi^0$. Crosses are data and histograms are PWA
                fit projections.}
    \label{gpp}
\end{figure}

The $\pi^+\pi^-$ and $\pi^0 \pi^0$ invariant mass distributions from
$J/\psi \to \gamma \pi^+\pi^-$ and $\gamma \pi^0 \pi^0$ are shown in
Fig. \ref{gpp} (left) and (right). The partial wave analyses are carried out
in the 1.0-2.3 GeV $\pi\pi$ mass range. Two $0^{++}$ states exist
in the mass lower than 1.8 GeV.
The first one peaks at $1466\pm6\pm16$ MeV with a width of 
$108^{+14}_{-11}\pm21$ MeV, which is approximately consistent 
with $f_0(1500)$. Due to the large interference between 
S-wave states, a possible contribution from $f_0(1370)$ cannot be 
excluded. The second $0^{++}$ peaks at around 1.75 GeV. If it is
the same state with that observed in $J/\psi\to\gamma K \bar{K}$~\cite{gkk},
we obtain the ratio of decaying to $\pi\pi$ and $K\bar K$ as $0.41^{+0.08}_{-0.15}$.
\vspace{-0.5cm}

\section{Study of the excited baryon states from $J/\psi \to p \bar n \pi^-$
+ $c.c$}

\begin{figure}[htpb]
\centerline{\psfig{file=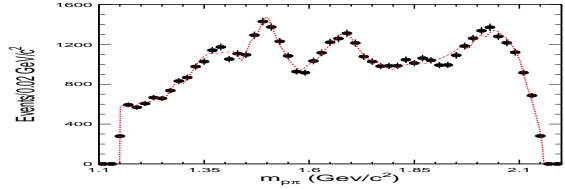,width=8.cm,height=2.9cm}}
 \caption{\it The invariant mass spectrum of $\pi N$}
    \label{invm} 
\end{figure}

The $\pi N$ system in decays of $J/\psi\to\bar NN\pi$ is limited
to be isospin 1/2 by isospin conservation. This provides a big
advantage in studying $N^*\to \pi N$ compared with $\pi N$ and
$\gamma N$ experiments which mix isospin 1/2 and 3/2 for the $\pi
N$ system. Fig. \ref{invm} shows the $\pi N$ invariant mass spectrum from 
$J/\psi \to p \bar n \pi^-$. Besides two well
known $N^*$ peaks at 1500 MeV and 1670 MeV, there are two new,
clear $N^*$ peaks in the $p\pi$ invariant mass spectrum around
1360 MeV and 2030 MeV.  They are the first direct observation of
the $N^*(1440)$ peak and a long-sought `missing" $N^*$ peak above
2 GeV in the $\pi N$ invariant mass spectrum.  A simple
Breit-Wigner fit gives the mass and width for the $N^*(1440)$ peak
as $1358\pm 6 \pm 16$ MeV and $179\pm 26\pm 50$ MeV, and for the
new $N^*$ peak above 2 GeV as $2068\pm 3^{+15}_{-40}$ MeV and
$165\pm 14\pm 40$ MeV, respectively.
\vspace{-1.1cm}

\section{Other Results}

The mass, width of $\eta_c$, as well as some of its decay branching
ratios are measured \cite{etac} with BESII $5.8 \times 10^7$ $J/\psi$
events. The $J/\psi \to \gamma f_2(1270) f_2(1270)$ decay is first 
observed and measured. We also present much improved
measurements on $J/\psi \to \pi^+ \pi^- \pi^0$ \cite{3pi}, 
$K_s K_L$ \cite{kskl} and $p \bar p$ \cite{ppb} decays.

The lepton flavor violation (LFV) is searched for from $J/\psi \to e \mu$, 
$\mu \tau$ and $e \tau$ decays, the LFV processes.
The observed signal events are consistent with the background level. The
upper limits of the decay branching fractions are set \cite{lep}.

We also search for the pentaquark state $\Theta(1540)$ in
$J/\psi \to K^0_SpK^- \bar n$ and $K^0_S\bar p K^+n$ final states with 
$K^0_S$ decaying to $\pi^+\pi^-$. No clear $\Theta$ signal is observed. 
The upper limits are set \cite{pentaq}.

\section*{Acknowledgments}
We acknowledge the staff of the BEPC and IHEP computing
center for their hard efforts.
This work is supported in part by the National Natural Science Foundation
of China under contracts Nos. 19991480, 10225524, 10225525, 10175060 (USTC),
and No. 10225522 (Tsinghua University), the Chinese
Academy of Sciences under contract No. KJ 95T-03, the 100 Talents Program 
of CAS under Contract Nos. U-11, U-24, U-25, and the Knowledge Innovation 
Project of CAS under Contract Nos. U-602, U-34 (IHEP); 
and by the Department
of Energy under Contract No. DE-FG03-94ER40833 (U Hawaii)

\end{document}